\documentclass[journal]{IEEEtran}
\usepackage{graphicx}
\usepackage[tight,footnotesize]{subfigure}

\begin{document}
%
\title{Construction and Performance of Large-Area Triple-GEM Prototypes for
  Future Upgrades of the CMS Forward Muon System\\
\vspace*{-5cm}{\tiny This work has been submitted to the IEEE
  Nucl. Sci. Symp. 2011 for publication in the conference record. Copyright
  may be transferred without notice, after which this version may no longer be
  available.}\\
\hspace*{15cm}{\bf\small RD51-Note-2011-012}\vspace*{3.2cm}}
%
%
%
\author{
M. Tytgat*,~\IEEEmembership{Member,~IEEE,} A.~Marinov, N. Zaganidis,
Y. Ban, J.~Cai, H.~Teng,
A.~Mohapatra, T.~Moulik,
M.~Abbrescia, A. Colaleo, G. de Robertis, F. Loddo, M.~Maggi, S.~Nuzzo,
S.~A.~Tupputi, 
L. Benussi, S. Bianco, S.~Colafranceschi, D. Piccolo, G. Raffone, G. Saviano,
M.G.~Bagliesi, R.~Cecchi, G. Magazzu, E. Oliveri, N.~Turini,
T.~Fruboes,
D.~Abbaneo, C.~Armagnaud, P. Aspell, S.~Bally, U. Berzano, J.~Bos, K.~Bunkowski,
J.~P.~Chatelain, J.~Christiansen, A.~Conde~Garcia, E. David, R.~De~Oliveira,
S.~Duarte Pinto, S. Ferry, F. Formenti, L.~Franconi, A.~Marchioro, K.~Mehta,
J.~Merlin, M.~V.~Nemallapudi,
H.~Postema,
A.~Rodrigues, L.~Ropelewski, A.~Sharma,~\IEEEmembership{Senior
  Member,~IEEE,} N. Smilkjovic, M.~Villa, M. Zientek,
A.~Gutierrez, P.~E.~Karchin,
K.~Gnanvo, M.~Hohlmann,~\IEEEmembership{Member,~IEEE,} M.~J.~Staib
\thanks{Manuscript received November 15, 2011}%
\thanks{A.~Marinov, M.~Tytgat, N.~Zaganidis are with the Dept. of
  Physics and Astronomy, Ghent University, Gent, Belgium}%
\thanks{Y.~Ban, J. Cai, H.~Teng are with Peking University, Beijing, China}%
\thanks{A.~Mohapatra, T.~Moulik are with NISER, Bhubaneswar, India}%
\thanks{M. Abbrescia, A. Colaleo, G. de Robertis, F. Loddo, M.~Maggi,
  S. Nuzzo, S.~A.~Tupputi are with 
        Politecnico di Bari, Universit\`a di Bari, and INFN Sezione di Bari,
        Bari, Italy}%
\thanks{L. Benussi, S. Bianco, S. Colafranceschi, D. Piccolo, G.~Raffone,
  G.~Saviano are with the
Labortori Nazionali di Frascati INFN, Frascati, Italy}%
\thanks{M.G.~Bagliesi, R.~Cecchi, G. Magazzu, E.~Oliveri, N.~Turini are with 
        INFN Sezione di Pisa, Pisa, Italy}%
\thanks{T.~Fruboes is with Warsaw University, Warsaw, Poland}%
\thanks{D.~Abbaneo, C.~Armagnaud, P. Aspell, S. Bally, U. Berzano, J.~Bos,
  K.~Bunkowski, 
  J.~P.~Chatelain, J.~Christiansen, A.~Conde~Garcia, E. David, R.~De~Oliveira,
  S.~Duarte Pinto, S. Ferry, F. Formenti, L. Franconi, A.~Marchioro,
K.~Mehta, J.~Merlin, M.V.~Nemallapudi,
  H. Postema, A. Rodrigues, L. Ropelewski, A. Sharma, N. Smilkjovic, M.~Villa,
  M. Zientek are with the
        Physics~Department,~CERN, Geneva,~Switzerland}%
\thanks{A. Gutierrez, P.~E.~Karchin are with the Dept. of Physics and Astronomy,
       Wayne State University, Detroit, MI, USA}%
\thanks{K. Gnanvo, M. Hohlmann, M.~J.~Staib are with the Dept. of Physics and
  Space Sciences,
        Florida Institute of Technology, Melbourne, FL, USA}%
\thanks{* Corresponding author, michael.tytgat@cern.ch}}

\maketitle
\pagestyle{empty}
\thispagestyle{empty}

\begin{abstract}
At present, part of the forward RPC muon system of the CMS detector at the
CERN LHC remains uninstrumented in the high-$\eta$ region. An international
collaboration is investigating the possibility of covering the
$1.6<|\eta|<2.4$ region of the muon endcaps with large-area triple-GEM
detectors. Given their good spatial resolution, high rate capability, and
radiation hardness, these micro-pattern gas detectors are an appealing option
for simultaneously enhancing muon tracking and triggering capabilities in a
future upgrade of the CMS detector. A general overview of this feasibility
study will be presented. The design and construction of small ($10\times
10$~cm$^2$) and full-size trapezoidal ($1\times 0.5$~m$^2$) triple-GEM prototypes will be described. During detector assembly, different techniques for stretching the GEM foils were tested. Results from measurements with x-rays and from test beam campaigns at the CERN SPS will be shown for the small and large prototypes. Preliminary simulation studies on the expected muon reconstruction and trigger performances of this proposed upgraded muon system will be reported.
\end{abstract}


%

\section{Introduction}

\begin{figure}[!ht]
\centering
\includegraphics[width=3.5in]{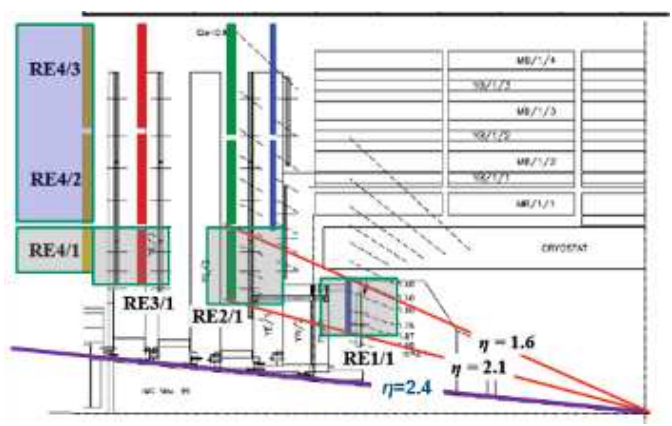}
\caption{One quarter of the CMS muon system. The colored detectors are the
RPCs in the endcap system. RPCs (REi/1 and RE4/2-3) enclosed in boxes have yet 
to be installed.}
\label{fig:cmsmuonsystem}
\end{figure}

\IEEEPARstart{T}{he} Compact Muon Solenoid (CMS) 
experiment~\cite{cmsdetpaper} has been collecting data successfully 
since the start of the first Large Hadron Collider (LHC) 
physics run in 2009. During two future long shutdown 
periods of the accelerator, the CMS Collaboration 
intends to upgrade several subsystems of its
detector~\cite{cmsupgradeTP}. 
In particular, the muon system as depicted in Fig.~\ref{fig:cmsmuonsystem}
will be
extended by completing rings 2 and 3 in the fourth station in both endcaps
to ensure efficient muon triggering and reconstruction in that
region as the LHC
instantaineous luminosity continues to increase. 
During the first Long Shutdown presently scheduled for
2013-2014, this fourth endcap station will be equipped with Resistive Plate
Chambers (RPCs) up to $|\eta|<1.6$. For the latter subsystem, which is a 
dedicated CMS muon trigger detector, the very forward region beyond
$|\eta|=1.6$ will remain empty and could in principle be instrumented up to
$|\eta|=2.4$ as is already the case for the Cathode Strip Chamber (CSC)
system that serves as muon tracker in the endcap region. However, the present 
design of the endcap RPCs, made of a double Bakelite gas gap and operating in
avalanche mode, is not expected to be suitable for the particle rates
amounting to several tens of kHz/cm$^2$ in the scenario of an LHC luminosity 
going up to $10^{34-35}$~cm$^{-2}$s$^{-1}$. 

Here, we report on an 
ongoing feasibility study to insert Gas Electron Multiplier
(GEM) detectors in the vacant space of the RPC endcap system beyond
$|\eta|=1.6$. In general, GEM detectors feature excellent spatial
($\sim$100~$\mu$m) 
and timing ($\sim$5~ns) 
resolution and are proven to be able to withstand particle rates up
to 10~MHz/cm$^2$. Furthermore, in the specific 
case of CMS, the use of such detector 
technology could in principle 
allow the combination of 
muon triggering and tracking capabilities into one single subsystem. 

Generally desired for a trigger/tracker 
detector in the CMS high $\eta$ region
are a time resolution better than 5~ns, a spatial resolution in the range
of 200-400~$\mu$m and an overall detector efficiency exceeding 97~\%.

\section{Studies with Small Prototypes}

%
\begin{table*}[!ht]
\renewcommand{\arraystretch}{1.3}
\caption{Overview of the small triple-GEM prototypes that were constructed and
  studied in this project.}
\label{tab:smallprototypes}
\centering
\begin{tabular}{|c|c|c|c|c|c|c|c|}
\hline
Name & Mask Type & Prod. & Active & Readout & Gap Sizes & \#strips & Prod.\\
& & Tech. & Area & & ({\it drift, transf. 1,
  transf. 2, ind.}) & & Site \\
& & & (cm$^2$) & & (mm/mm/mm/mm) & &\\
\hline
"Timing GEM" & double-mask & standard & $10\times 10$ & 1D & (3/2/2/2)\&(3/1/2/1) & 128 & CERN \\
"Single-Mask GEM" & single-mask & standard & $10\times 10$ & 2D & (3/2/2/2) & 512 &
CERN \\
"Honeycomb GEM" & double-mask & standard & $10\times 10$ & 2D & (3/2/2/2) & 512 & CERN
\\
"CMS Proto III" & single-mask & self-stretching & $10\times 10$ & 1D & (3/1/2/1) & 256 & CERN \\
"CMS Proto IV" & single-mask & self-stretching & $30\times 30$ & 1D & (3/1/2/1) & 256 & CERN \\
"Korean I" & double-mask & standard & $8\times 8$ & 1D & (3/2/2/2) & 256 & New Flex\\
\hline
\end{tabular}
\end{table*}

 
In the course of this project that was initiated in 2009, several different
types of small triple-GEM prototypes were produced and then 
studied using x-rays 
in the
RD51~\cite{rd51} lab of the CERN Detector Technology Group (DT) 
and/or
particle beams at the CERN SPS. Table~\ref{tab:smallprototypes} specifies
the small 
detectors
that were produced so far along with their main construction parameters.
The first three small triple-GEMs listed in the table were extensively 
tested
during 
2009-2010 using the
150~GeV pion/muon beam of the CERN SPS H4 beam line. The main
test results are summarized below; more details can be found in~\cite{CMSGEM1}.

The "Timing
GEM" was mainly used to study the time resolution that could be obtained
with such detectors, as function of the used gas mixture, and the drift and
induction fields. A time resolution of 4~ns could be reached with an
Ar/CO$_2$/CF$_4$ 45/40/15 gas mixture and a 3/1/2/1~mm (drift, tranf.1, 
transf.2, ind.) 
gap size configuration, which meets the requirement for the CMS muon
triggering system.   

Adopting a geometry for the GEMs similar to the RPCs in the CMS endcap disks,
the smallest active GEM detector area needed by CMS is of the order 
of 50x100~cm$^2$. 
For such sizes, the standard double-mask technique to produce the GEM foils 
is not ideal as it suffers from alignment problems of the two masks on either 
side of the foils during the photolitographic hole etching process. 
The single-mask technique~\cite{singlemask} 
overcomes this problem and was used to produce the 
"Single-Mask GEM" prototype. The performance of
the single-mask GEM was quite similar to our ``Timing GEM''. 
An efficiency up to 98~\% was
measured, albeit for a slightly higher gain than for the double-mask GEM. 
Nevertheless, the single-mask technique appears quite mature and was chosen for
the production of our large prototypes.

Depending on how many endcap disks would be instrumented with GEMs, the
number of detectors needed for CMS could amount to several 100s. For such
quantities, the time and certainly the cost of the detector production becomes 
an issue. The most time-consuming and labor-intensive part of triple-GEM
production is the foil stretching and the gluing of the spacer frames.  
To avoid these steps in the production process, two novel assembly procedures
were tested as described below.

The "Honeycomb GEM" produced in 2010, was assembled using honeycomb structures
as spacers in the detector gaps between the GEM foils, which avoids the need
to stretch the foils. Although this prototype could be operated without any
problems, very
localized efficiency losses were observed at the position of the honeycomb
structures. With honeycomb cell sizes of (6/12/12/12)mm or (6/0/0/0)mm 
in the (drift, transf.1, transf.2, ind.) gap, an overall detector efficiency
of about 75~\% was obtained.  

\begin{figure}[!ht]
\centering
\subfigure[]{\includegraphics[width=1.7in]{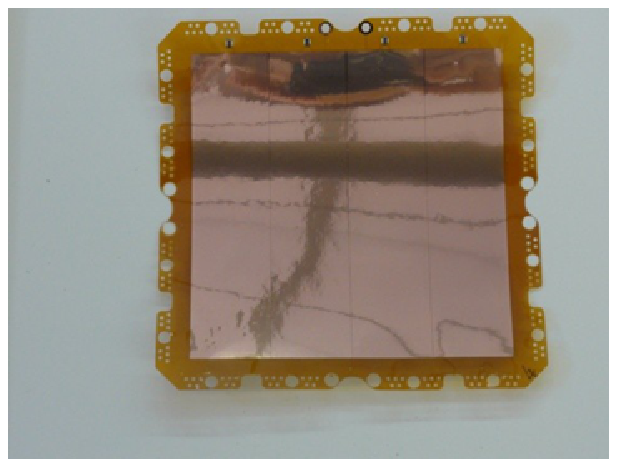}\label{fig:ns2a}}
\hfil
\subfigure[]{\includegraphics[width=1.7in]{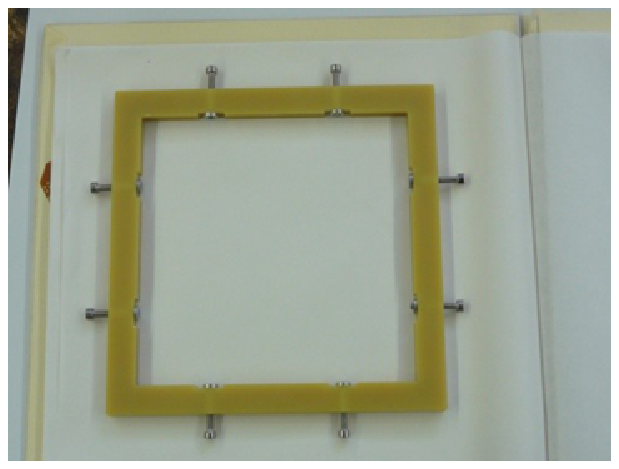}\label{fig:ns2b}}

\subfigure[]{\includegraphics[width=1.7in]{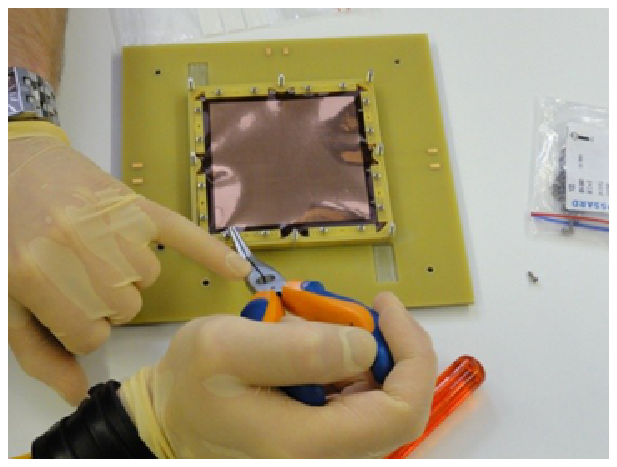}\label{fig:ns2c}}
\hfil
\subfigure[]{\includegraphics[width=1.7in]{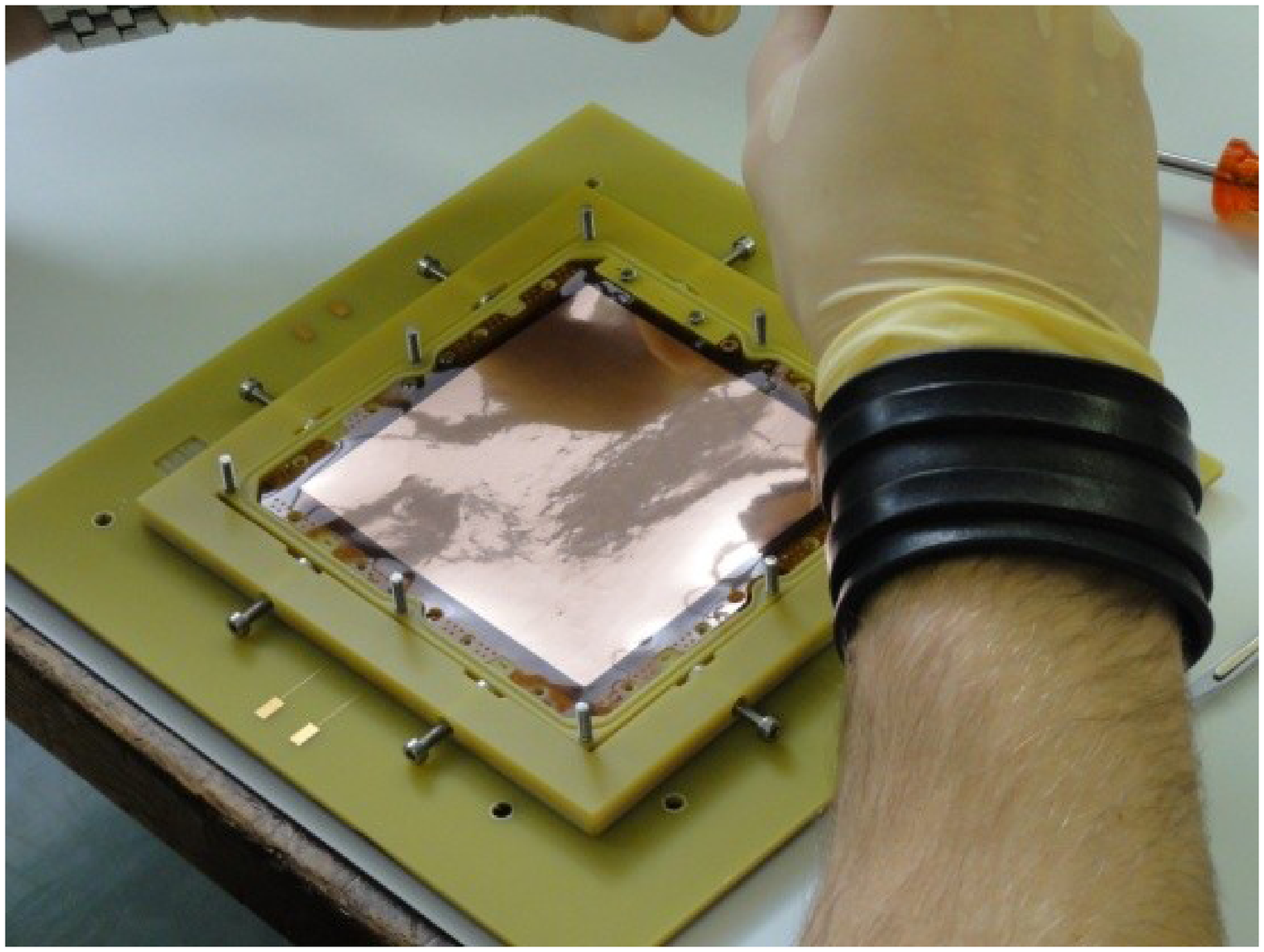}\label{fig:ns2d}}

\subfigure[]{\includegraphics[width=1.7in]{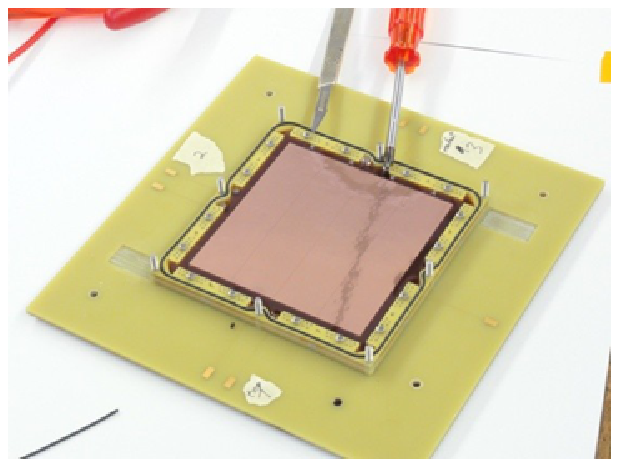}\label{fig:ns2e}}
\hfil
\subfigure[]{\includegraphics[width=1.7in]{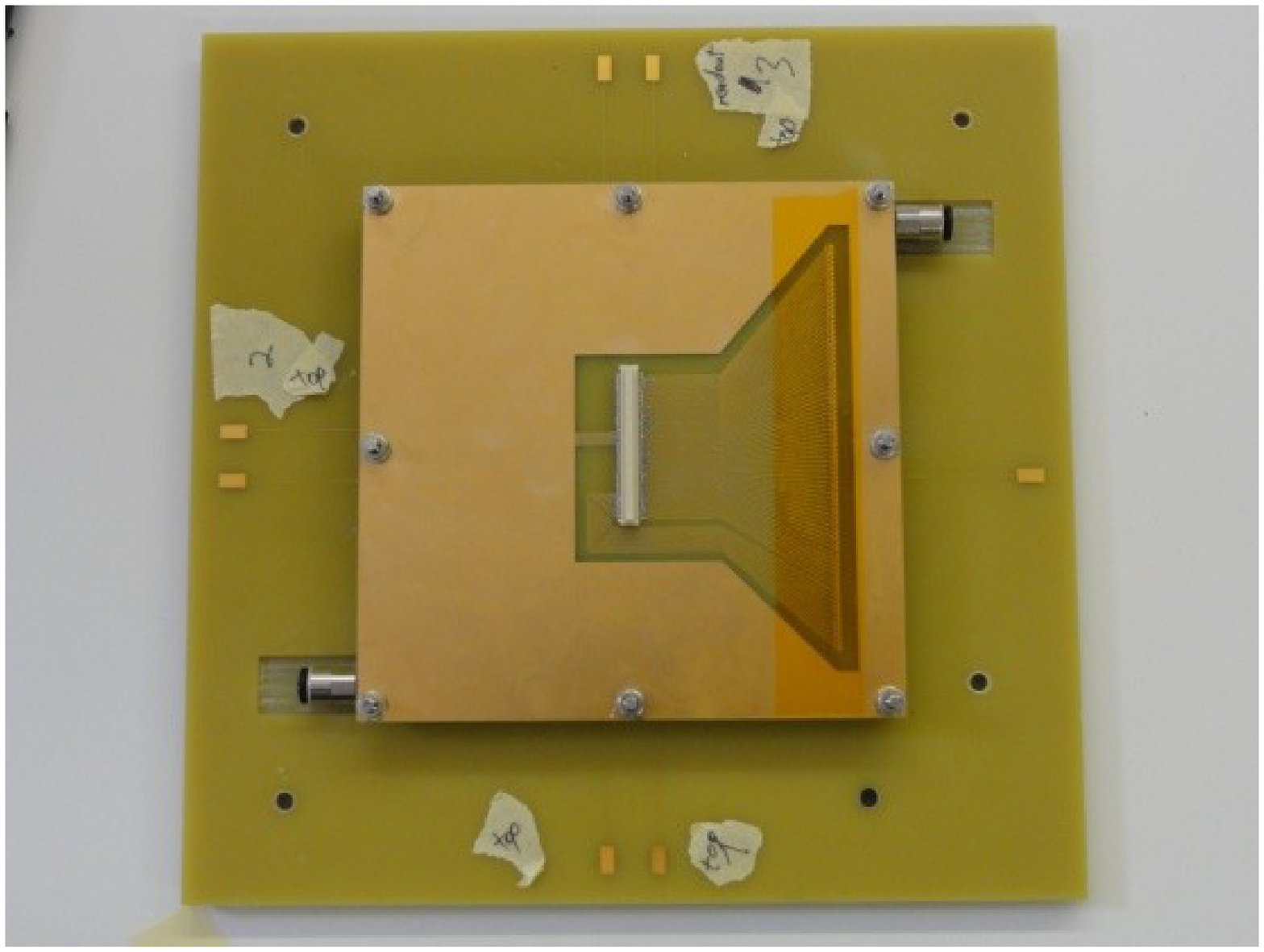}\label{fig:ns2f}}
\caption{The {\it self-stretching} triple-GEM assembly technique : (a) GEM
  foil before assembly; (b) external frame for foil stretching; 
(c) mounting the GEM
  foils; (d) stretching the foils; (e) stretched GEM; (f) completed detector
  including readout board.}
\label{fig:ss}
\end{figure}

An important development in 2011 was the introduction of another new GEM
assembly technique, here referred to as the {\it self-stretching}
technique. The procedure is demonstrated in Fig.~\ref{fig:ss} showing 
a few photographs taken
during the assembly at CERN 
of the first $10\times 10$~cm$^2$ triple-GEM, ``CMS Proto III'', 
prototype using this new technique.
The GEM foils are produced with a pattern of holes in the edge on each of 
the four sides. The foils
are then placed on a detector structure which has a set of alignment pins 
matching the
hole pattern in the foil edges. Next, using a special external frame that 
is placed around 
the ensemble, the foils are mechanically stretched and then fixed with 
screws to the detector structure. In the end, once the foils have been
completely stretched and fixed, the readout board can be mounted, closing 
the detector.

Clearly, compared to the standard assembly procedure, the 
{\it self-stretching} technique
offers many advantages. No gluing nor soldering is
required during the assembly procedure and the detector is produced without
the need to place any spacers in the active area, in the gaps between the 
foils. The technique is very fast, for example this small prototype
was assembled in only 1 hour. As an additional benefit, 
if needed, it allows for the 
detector to be re-opened in order to make modifications or repairs, or to 
replace a GEM foil. 

The small ``CMS Proto III'' prototype was tested using an x-ray gun with a Cu
target 
in the RD51 lab. With an Ar/CO$_2$ 70/30 gas mixture, the detector exhibited 
stable operation for a measured gain up
to at least $3\cdot 10^4$. The detector response was also
observed to be very uniform across the GEM surface. 
Given these promising results,
another prototype with dimensions 
$30\times 30$~cm$^2$ was just recently (Sept. 2011) 
produced with the new technique. Initial tests of its performance are ongoing.

\section{Studies with Large Prototypes}

\begin{figure}[!ht]
\centering
\subfigure[]{\includegraphics[width=3in]{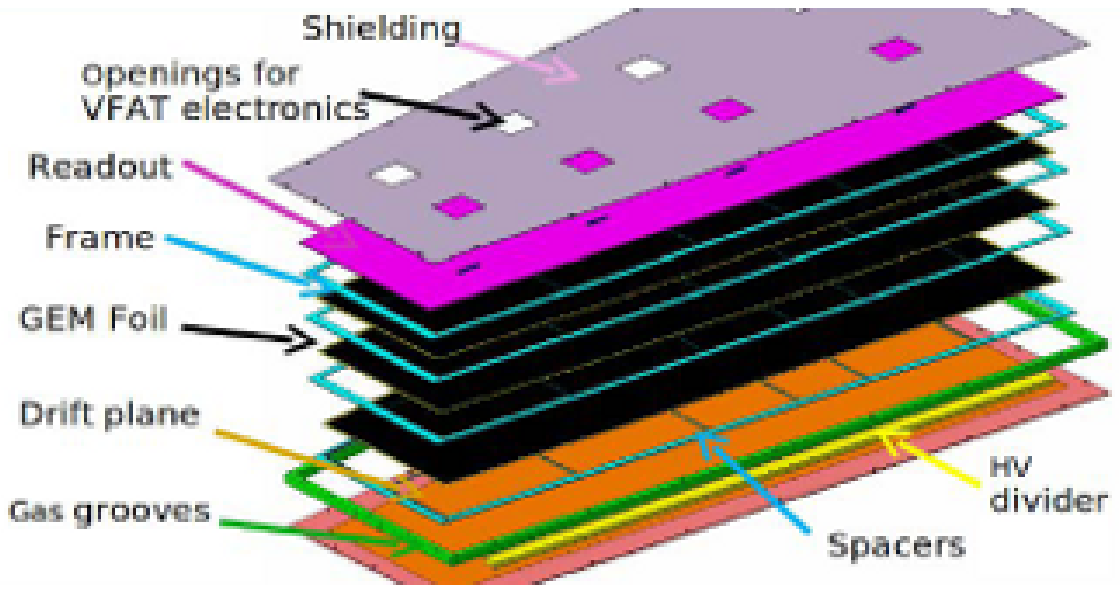}
\label{fig:GE11diagram_a}}
\subfigure[]{\includegraphics[width=3.5in]{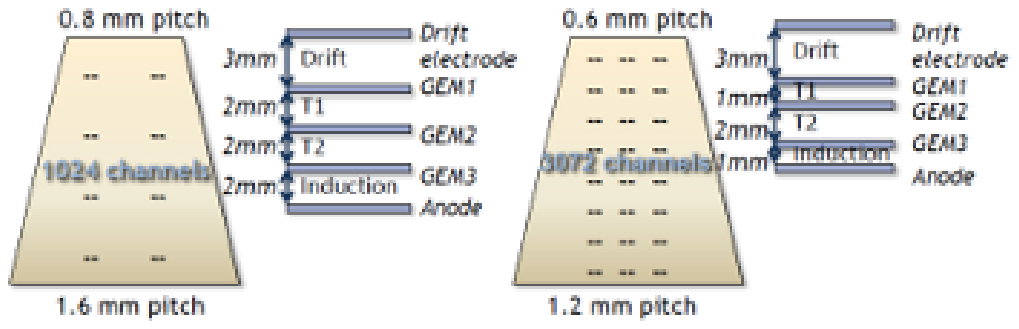}
\label{fig:GE11diagram_b}}
\caption{(a) Exploded view of the first full-size triple-GEM prototype for
  CMS. (b) Layout and gap size configuration of both full-size prototypes.}
\label{fig:GE11diagram}
\end{figure}

Besides the small prototypes described above, two full-size triple-GEM
prototypes were produced, called ``CMS Proto I'' (April 2010) and ``CMS Proto
II'' (March 2011).   
The design of the first prototype was reported earlier
in~\cite{CMSGEM2}. 
The large prototypes have a trapezoidal shape with a GEM active area of
$990\times (220-455)$~mm$^2$. Their geometry follows the
design that was foreseen for the RPCs to be installed in the first endcap 
disks in the high-$\eta$ region, ie. RE1/1 in Fig.~\ref{fig:cmsmuonsystem}. 
In fact, the GEM detectors are embedded in a similar aluminum cover box as 
was designed for the RPCs in that area. The chambers each cover a 
10$^{\circ}$ sector of the endcap disks. Fig.~\ref{fig:GE11diagram_a} shows an
exploded view of ``CMS Proto I'', while Fig.~\ref{fig:GE11diagram_b} displays
the general layout and gap size configuration of both detectors.

\begin{figure}[!ht]
\centering
\subfigure[]{\includegraphics[height=1.22in]{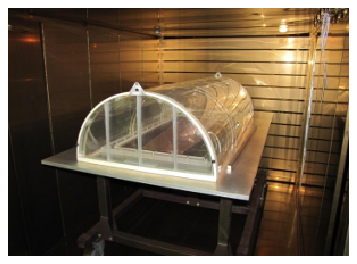}}
\hfill
\subfigure[]{\includegraphics[height=1.22in]{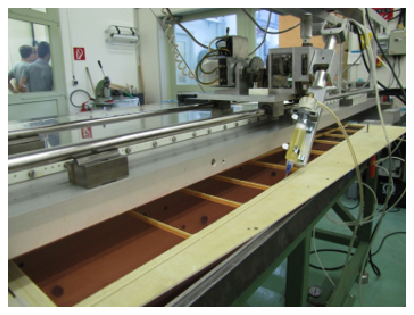}}

\subfigure[]{\includegraphics[height=1.3in]{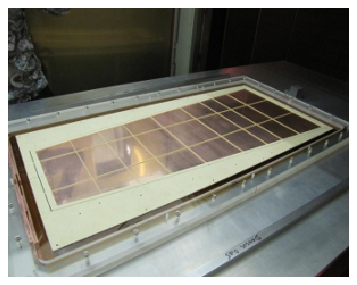}}
\hfill
\subfigure[]{\includegraphics[height=1.3in]{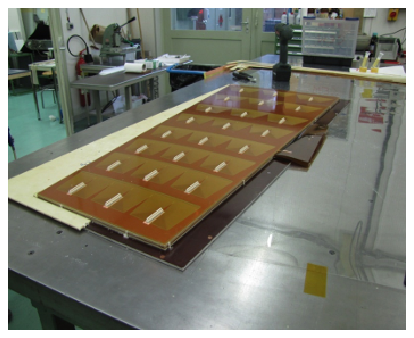}}
\caption{Assembly procedure of the full-size CMS prototypes : (a) foil
  stretching in the oven; (b) gluing the spacer frames; (c) curing the glue;
  (d) mounting the readout plane.} 
\label{fig:largeprotoassembl}
\end{figure}

\begin{figure}[!ht]
\centering
\includegraphics[width=2.5in]{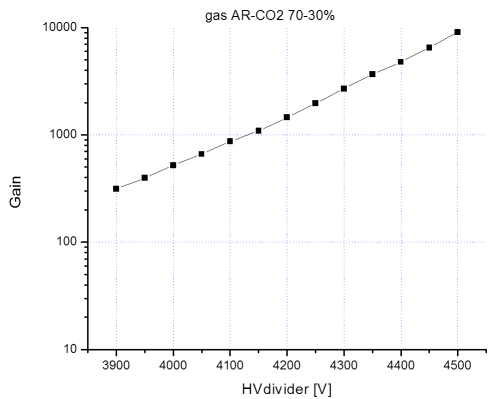}
\caption{Gain curve for the ``CMS Proto I'' measured with x-rays in the RD51 lab.}
\label{fig:GE11_gain}
\end{figure}

The GEM foils for both large-area prototypes were produced at CERN using a
photolithographic process with a 
50~$\mu$m thick kapton sheet with a 5~$\mu$m copper cladding on both sides.
Given the size of these large prototypes, 
the single-mask technique had to be used to produce the foils. To limit the
discharge probability, the foils are divided in 35 sectors of about 
100~cm$^2$ each. 
The drift electrodes, made of a 300~$\mu$m kapton foil with a 5~$\mu$m copper
cladding, were glued directly to the 3~mm thick
bottom aluminum plate of the detector
cover boxes. The readout planes provide a one-dimensional readout with 
radial strips,
i.e. pointing to the LHC beam line, with a pitch varying from 0.8 (short side)
to 1.6~mm (long side) for ``CMS Proto I'' and from 0.6 to 1.2~mm for ``CMS
Proto II''. The first prototype was divided into four $\eta$-partitions with 
256 strips per partition, while
the second had eight partitions with 384 strips each. 
Following the results obtained with the small prototypes on the GEM timing
performance, the gap size 
configuration for the second large prototype was 3/1/2/1~mm (drift,
transf.1, transf.2, induction) in contrast to the 3/2/2/2~mm configuration
for the first prototype.
Different steps of the assembly procedure of the large prototypes are 
depicted in Fig.~\ref{fig:largeprotoassembl}. Fig.~\ref{fig:GE11_gain} shows
the gain curve for ``CMS Proto I'', measured with an x-ray gun in the RD51 lab.

\begin{figure}[!ht]
\centering
\subfigure[]{\includegraphics[width=1.7in]{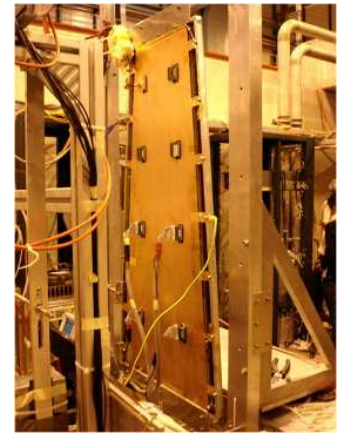}
\label{fig:cmslargeprototb_a}}
\hfill
\subfigure[]{\includegraphics[width=2.5in]{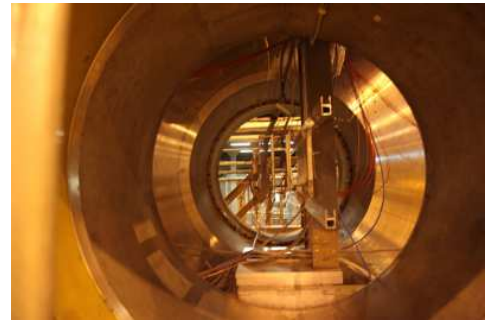}
\label{fig:cmslargeprototb_b}}
\caption{(a) The ``CMS Proto I'' detector installed at
  the SPS H4 beam line. (b)~The ``CMS Proto II''
  detector installed inside the CMS M1 magnet at the SPS H2 beam line.} 
\label{fig:cmslargeprototb}
\end{figure}

The large prototypes were tested in 150~GeV 
pion/muon beams at the CERN SPS H2 and H4
beamlines
during several campaigns in October 2010, and from June to September 
2011. Fig.~\ref{fig:cmslargeprototb_a} shows the first large-area prototype
installed at the SPS H4 beam line in October 2010. A special mounting frame
allowed to move the
detector in the transverse plane with respect to the beamline,
such that the beam could be pointed at different locations along the GEM
surface in order to study the uniformity of the detector response.
The influence of a magnetic field on the performance
of the prototypes was also tested by inserting them in the M1 
magnet at the SPS H2 beamline. This magnet can produce 
fields up to 3~T.
Inside the CMS detector, the GEMs would be
installed at a location where the magnetic field can reach up to 1.5~T with 
an angle between the magnetic field and the GEM electric field
less than 8$^{\circ}$. 

\begin{figure}[!ht]
\centering
\subfigure[]{\includegraphics[width=2.2in]{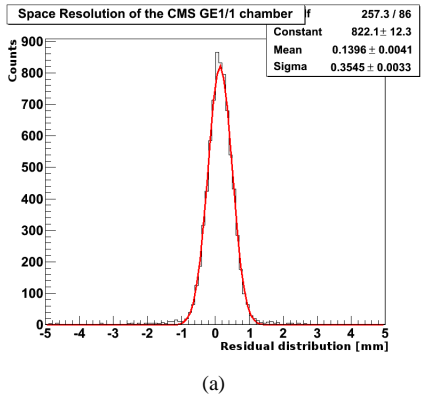}
\label{fig:GE11_I_results_a}}

\subfigure[]{\includegraphics[width=2in]{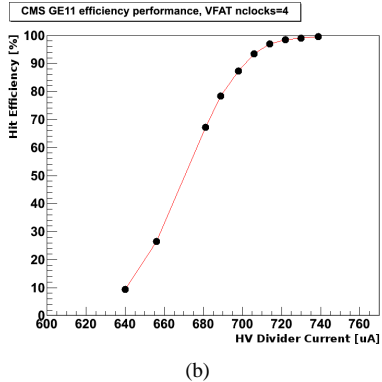}
\label{fig:GE11_I_results_b}}
\caption{Results obtained with ``CMS Proto I'' during the Oct. 2011 test beam
  campaign : (a) residual distribution; (b) detector hit efficiency as
  function of the current in the HV divider (proportional to the electric 
fields in the GEM).}
\label{fig:GE11_I_results}
\end{figure}

Fig.~\ref{fig:GE11_I_results} shows a few results obtained with ``CMS Proto
I'' in 
the October 2010 campaign. Using an Ar/CO$_2$ 70/30 gas mixture, with
the beam pointing at the detector $\eta$-sector with the smallest average 
pitch size, ie. 1.05~mm, a 
spatial resolution near 
the value expected from the strip pitch ($1050/\sqrt{12}=305$~$\mu$m), and 
a detector efficiency above 98~\% were obtained.
The uniformity of the
detector response across the GEM active area was found to be very good. 

\begin{figure}[!ht]
\centering
\subfigure[]{\includegraphics[width=2.1in]{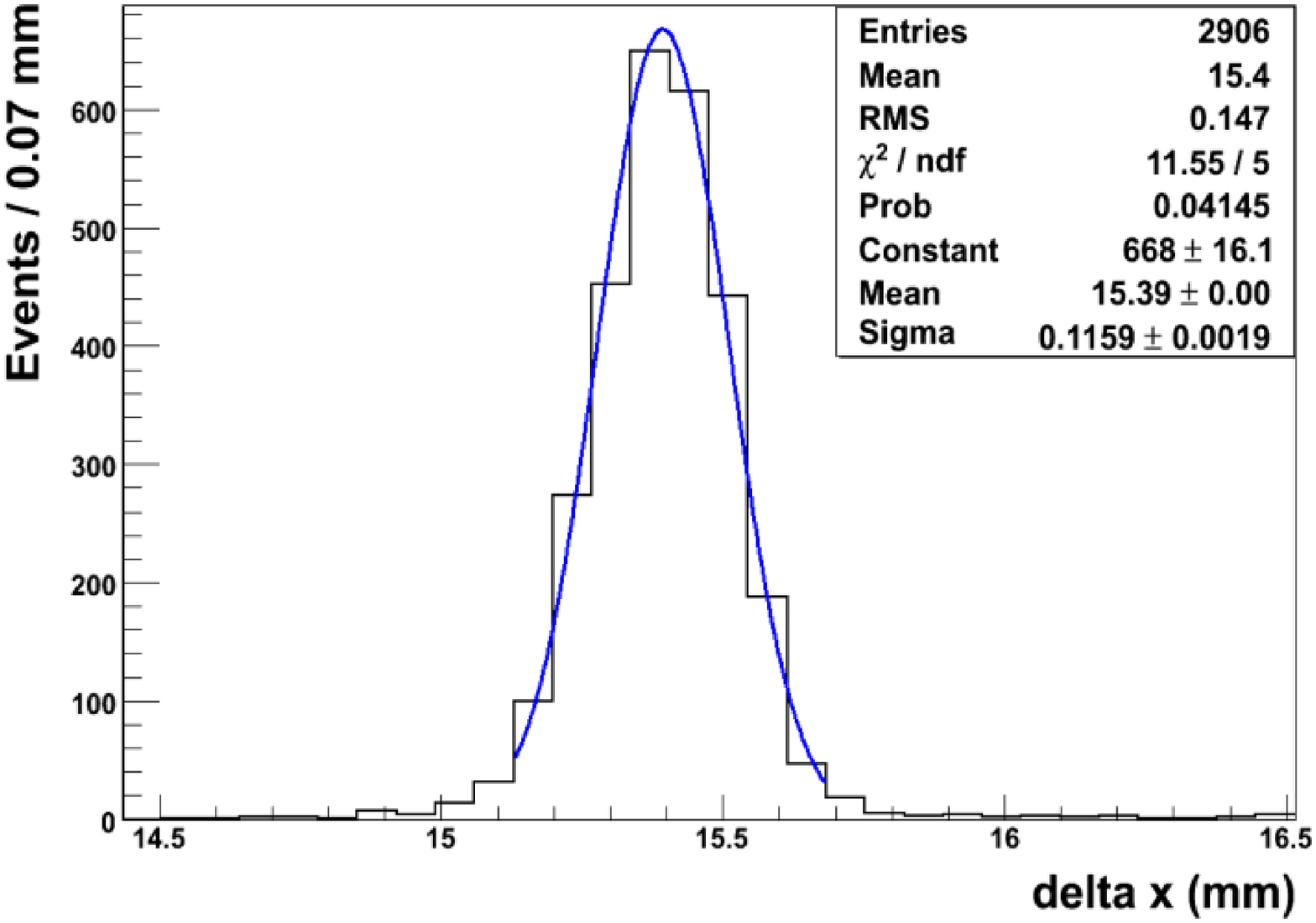}}

\subfigure[]{\includegraphics[width=2.1in]{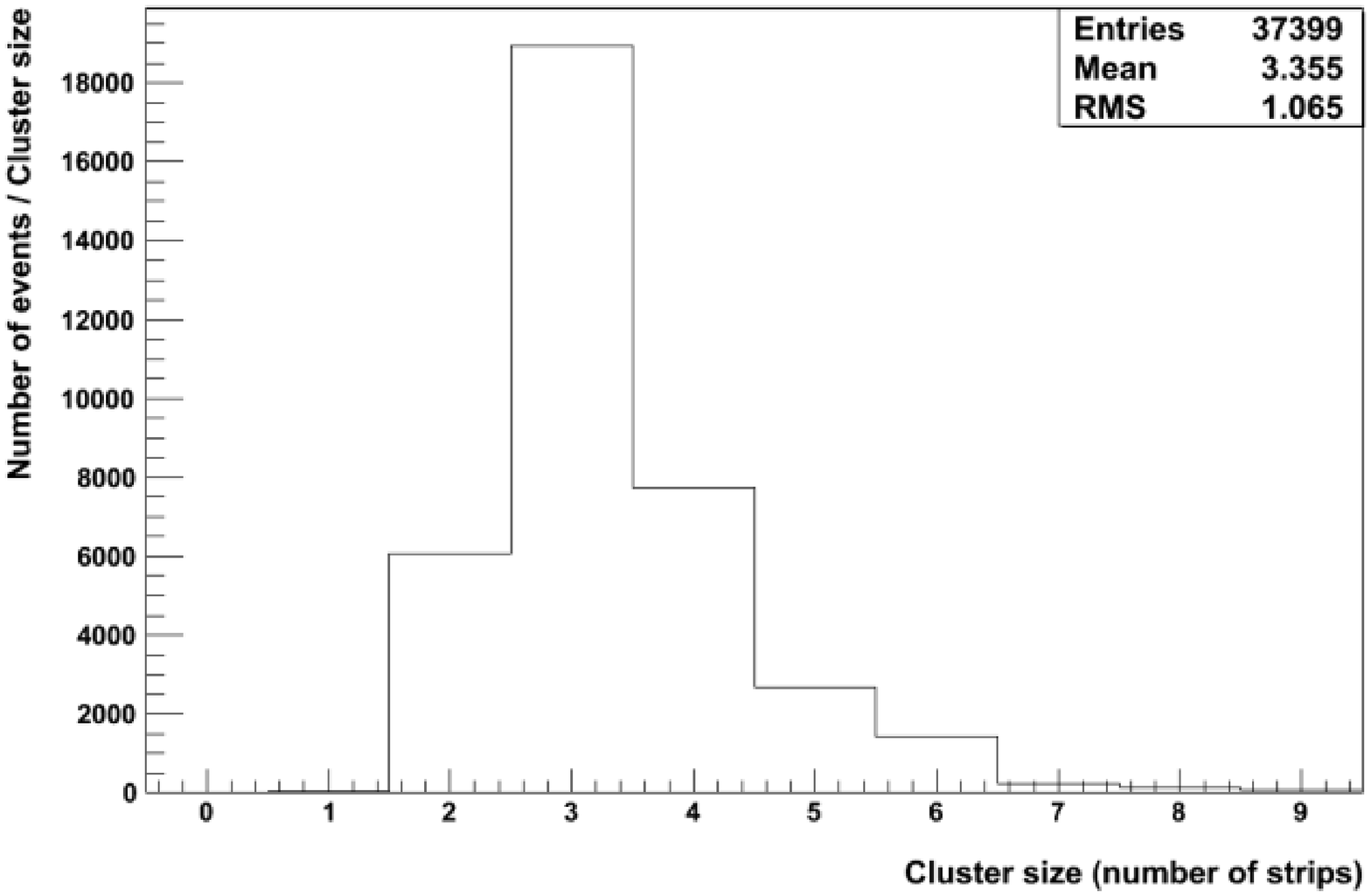}}
\caption{Results obtained with ``CMS Proto II'' during the Aug. 2011 test
  beam campaign : (a) difference $\Delta x$ between hit positions measured with
  ``CMS Proto II'' and with a reference tracker GEM; (b) strip cluster
size distribution.}
\label{fig:GE11_II_SRS}
\end{figure}

In Fig.~\ref{fig:GE11_II_SRS} some results are shown for the ``CMS Proto II'' 
detector, which has a smaller strip pitch than ``CMS Proto I'', 
obtained during the August 2011 test beam campaign. With the
detector flowing an 
Ar/CO$_2$/CF$_4$ 45/15/40 gas mixture, the data were
taken with the H4 pion beam pointing at the region with the 
smallest strip pitch, i.e. 573~$\mu$m. An average cluster size of about 3 strips
was found and for the space resolution an upper limit of 103~$\mu$m could be
derived using the information on the charge sharing among adjacent strips,
which is significantly better than the resolution expected from 
a purely digital readout (as used for the result in
Fig.~\ref{fig:GE11_I_results_a}), i.e. 
$strip~pitch/\sqrt{12}=573/\sqrt{12}=165$~$\mu$m. 

\begin{figure}[!ht]
\centering
\includegraphics[width=2in]{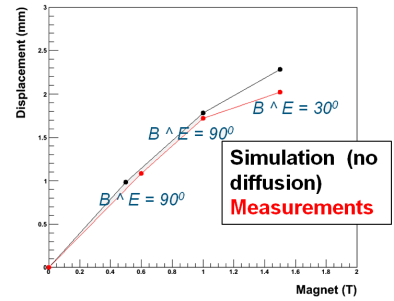}
\caption{Comparison of measured and simulated strip cluster position 
displacement in ``CMS Proto II'' 
due to an external magnetic field.}
\label{fig:GE11_M1}
\end{figure}

In Fig.~\ref{fig:GE11_M1} the measured shift in the average strip
cluster position due to a magnetic field is
shown for ``CMS Proto II'' inside the M1 magnet. Measurements were performed
in June and July 2011 using the SPS H2 150~GeV muon beam, with the detector 
flowing an
Ar/CO$_2$/CF$_4$ 45/15/40 gas mixture. The angle between the GEM electric
field and the external magnetic field could be varied by rotating the detector
inside the M1 magnet.
The data are in good agreement with a
Garfield/Magboltz~\cite{garfield-magboltz} simulation performed for
a 90$^{\circ}$ angle between the magnetic field and the GEM electric
field. The measurement at 1.5~T was performed for an angle of 30$^{\circ}$,
which explains the observed small deviation between data and simulation for
that point.

Up until recently, 
the readout of both the small and large 
prototypes was 
based on the TURBO front-end electronics with the VFAT~\cite{VFAT} chip that 
produces digital output for its 128 
analogue input channels. Efforts are ongoing to design a new, enhanced version
of the VFAT2 chip to replace the present system. Furthermore, in the
August-September 
2011
campaigns the prototypes were for the first time successfully operated with 
the APV25
chip in combination with the Scalable Readout System (SRS)~\cite{RD51SRS} 
that is being developed by the RD51 Collaboration. The results shown in
Fig.~\ref{fig:GE11_II_SRS} were obtained with this system.

More details on test beam results obtained with the two large-area prototypes 
can be found in~\cite{largeprototbresults}.

\section{GEM Production Facilities}

One of the possible scenarios that is presently under investigation for the
installation of GEMs in the CMS detector, proposes the use of double chambers
in each endcap disk, i.e. two chambers installed face-to-face. This would
double
the total number of detection layers and would thus 
greatly improve the tracking 
capability of the system. In such a scenario, to equip     
the $1.6<|\eta|<2.4$ region of the first and second CMS endcap disks, ie. the 
disks closest to the beam interaction point on either side, 
a total GEM foil area of order 500~m$^2$ would be needed. In such a case, 
an industrialized production process becomes imperative. 

With the ever-increasing demand for GEMs at CERN, the production facility will
be significantly expanded in 2011-2013. 
A completely new production line with machines suitable for large numbers
of large-area GEMs, up to about $2\times 0.6$~m$^2$, 
is presently being put into place.
In the end, the new facility
will be housed in a dedicated building (\#107) at CERN.
With this new setup, it is 
expected that
the production cost of large-area GEMs could be lowered to about 1~kCHF/GEM, 
with a production rate of 240 GEMs/year. 

Recently, an alternative GEM production facility was put in place at the New
Flex company near Seoul in South Korea. Initial contacts with this company
were made by CERN in November 2008 and in June 2011 a full technology
transfer from CERN to New Flex was organized. In the meantime, the company has
produced its very first $8\times 8$~cm$^2$ double-mask triple-GEM, complete with 
voltage divider.
The ``Korean I'' prototype (see Table~\ref{tab:smallprototypes}) was
tested
with a Cu target x-ray gun in the RD51 lab. The detector and voltage 
divider exhibited very stable operation with an Ar/CO$_2$ 70/30 gas mixture; 
smooth exponential gain curves were measured, similar to the CERN GEMs. The
company is now progressing towards the production of large-area GEMs.

\begin{figure}[!ht]
\centering
\subfigure[]{\includegraphics[width=3.5in]{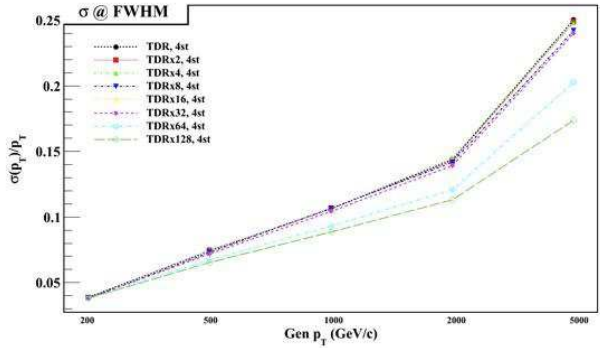}
\label{fig:triggersim_a}}

\subfigure[]{\includegraphics[width=3.5in]{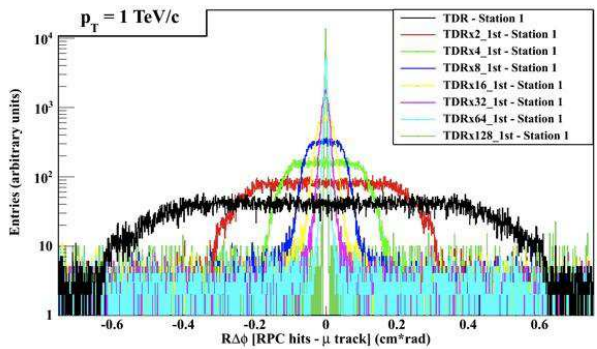}
\label{fig:triggersim_b}}
\caption{Monte Carlo simulations for various detector granularities of the
  extended CMS muon endcap system : (a) muon momentum resolutions; (b) simulated spatial residual distributions for $p_T=1$~TeV muons.}
\label{fig:triggersim}
\end{figure}

\section{Simulation Studies}

The CMS Monte Carlo simulation package is used to study and quantify 
the improvements in the trigger and muon reconstruction performance of an 
extended muon system in the CMS forward region as proposed here.
As a starting point the geometry of the RPCs as described in the CMS Muon
Technical Design Report~\cite{cmsmuondetector} is taken ($TDR$). To mimic the
GEM spatial resolution, the strip widths of the RPCs in the forward region
$1.6<|\eta|<2.1$ are gradually increased ($n\times TDR$) for 
the first two endcap 
stations. Using single-muon Monte Carlo simulation 
samples the effect of the GEM 
spatial resolution on the muon reconstruction can be studied. Examples of first
results on the reconstructed muon momentum resolution and track 
residuals are
shown in Fig.~\ref{fig:triggersim} for various simulated GEM spatial
resolutions. For a $p_T=1$~TeV muon, up to a strip segmentation a factor 8 
higher than in the current RPCs, the tracking resolution is dominated by the
strip width; for even smaller strip widths the residual distribution becomes 
gaussian.
These initial simulations are now being extended to study the 
improvements in the CMS muon trigger system. The effect of the extended muon
system on the detector acceptance for various physics channels under study
at CMS is also being addressed.

\begin{figure}[!ht]
\centering
\subfigure[]{\includegraphics[width=2.1in]{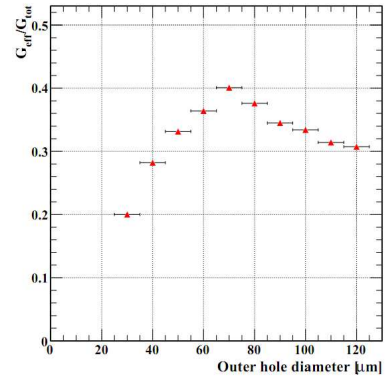}
\label{fig:gemsim_a}}

\subfigure[]{\includegraphics[width=2.1in]{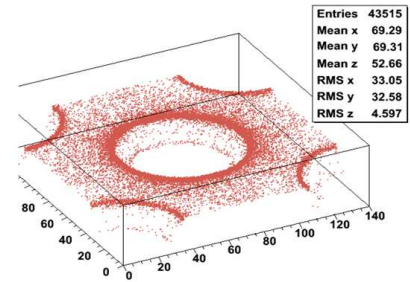}
\label{fig:gemsim_b}}
\caption{Example results from GEM simulation studies~: (a) effect of the outer
hole diameter on the GEM gain; (b) locations of avalanche electron losses in the GEM 
material near a GEM hole.}
\label{fig:gemsim}
\end{figure}

Simulations of the GEM 
detectors themselves are also being pursued~\cite{taniagemsim}. 
The modeling of the GEM is 
based on 
ANSYS, while the Garfield/Magboltz packages are used to simulate the 
electron avalanches and interactions with the material. With these simulations
the GEM gain is studied, and possible influences on the gain from various 
construction parameters, such as the outer hole 
diameter (see Fig.~\ref{fig:gemsim_a}), the drift and induction field,  
the thickness of the metal layers,
and also the Penning effect for the Ar/CO$_2$ gas mixture are addressed.
Detailed studies of the electron avalanches also allow to study electron
losses due to the GEM geometry or attachment to gas molecules. As an example,
Fig.~\ref{fig:gemsim_b} shows the position coordinates of electrons lost in
the GEM material. Finally, the
effect of an external magnetic field on the GEM performance is also studied
in detail.
   
\section{Summary and Outlook}

After intense studies on small-size prototypes, two fully
operational large-area triple-GEM detectors were designed and 
assembled during 2010-2011. With several test beam campaigns at 
the CERN SPS H2 and H4 muon/pion beam lines it was demonstrated that these
$990\times (445-220)$~mm$^2$ 
prototypes exhibit stable and reliable operation and can be run with 
high gain and detector efficiency. The prototypes were tested inside
a 1.5~T magnetic field, and could also be operated successfully with the RD51
Scalable Readout System. GEM detectors based on these full-size prototypes 
are being proposed for the
extension of the CMS muon endcap system in the forward region. Simulation
studies of the improved CMS trigger and physics 
performance with this system are ongoing.

\section*{Acknowledgment}
This work has partly been performed in the framework of the RD51
Collaboration.
The authors that are members of the CMS Collaboration would like to 
explicitly thank the 
RD51 Collaboration  
for its continuous support of this work, for the many
fruitful discussions, for the close collaboration during the test beam
campaigns and for dedicating part of their lab space at CERN to this project.

\ifCLASSOPTIONcaptionsoff
  \newpage
\fi


\begin{thebibliography}{1}

\bibitem{cmsdetpaper} S. Chatrchyan {\it et al.}, The CMS Collaboration, \emph{The CMS experiment at the CERN LHC}, J. Instrum. 3
  (2008) S08004
\bibitem{cmsupgradeTP} The CMS Collaboration, \emph{Technical Proposal for the
  Upgrade of the CMS detector through 2020}, CERN-LHCC-2011-006; CMS-UG-TP-1; LHCC-P-004
\bibitem{rd51}The RD51 Collaboration,
  http://rd51-public.web.cern.ch/RD51-Public/ 
\bibitem{CMSGEM1}D. Abbaneo {\it et al.}, \emph{Characterization of GEM
  Detectors for Application in the CMS Muon Detection System}, 2010 IEEE
  Nucl. Sci. Symp. Conf. Rec 1416-1422; RD51 Note 2010-005; arXiv:1012.3675v1 [physics.ins-det]
\bibitem{singlemask} S. Duarte Pinto {\it et al.}, JINST 4 (2009) P12009;
  M. Villa {\it et al.}, Nucl. Instr. and Meth. A628 (2011) 182-186
\bibitem{CMSGEM2}D. Abbaneo {\it et al.}, \emph{Construction of the first
  full-size GEM-based Prototype for the CMS High-$\eta$ Muon System}, 2010
  IEEE Nucl. Sci. Symp. Conf. Rec. 1909-1913; RD51 Note 2010-008;
  arXiv:1012.1524v2 [physics.ins-det]
\bibitem{garfield-magboltz}R.~Veenhof, {\it Garfield - simulation of gaseous
  detectors}, http://garfield.web.cern.ch/garfield/; S.~Biagi, {\it Magboltz -
  transport of electrons in gas mixtures}, http://consult.cern.ch/writeups/magboltz/
\bibitem{VFAT}  P. Aspell {\it et al.}, {\it VFAT2 : A front-end system on chip providing fast
trigger information, digitized data storage and formatting
for the charge sensitive readout of multi-channel silicon
and gas particle detectors}, Proc. of the Topical Workhop on Electronics for
  Particle Physic (TWEPP2007), Prague, Czech Republic, September 3-7, 2007
\bibitem{RD51SRS}S. Martoiu {\it et al.}, {\it Front-End Electronics for the
  Scalable Readout System of RD51}, Contribution N43-5 to this conference
\bibitem{largeprototbresults}D. Abbaneo {\it et al.}, {\it Test Beam
  Results of the GE1/1 Prototype for CMS High-Eta Muon System Future Upgrade},
  Contribution NP5.S-221 to this conference
\bibitem{cmsmuondetector} The CMS Collaboration, {\it The Muon Project, CMS
  Technical Design Report}, CERN/LHCC 97-32; CMS-TDR-003
\bibitem{taniagemsim} T. Moulik, {\it Simulation of a Triple-GEM detector for
  a potential CMS muon tracking and trigger upgrade}, Submitted to the 
Proc. of the 
  Technology and Instrumentation in Particle Physics 2011 (TIPP 2011) 
Conference, 9-14 June, 2011, Chicago, IL, USA
\end{thebibliography}
\end{document}